\documentclass{aa}  
\usepackage{graphicx}
\usepackage{mathtools}
\usepackage{amsmath}
\usepackage{fancyref}
\usepackage{longtable}
\usepackage{rotating}
\usepackage{pdflscape}
\usepackage{enumerate}
\usepackage[citecolor=blue, urlcolor=blue, linkcolor=blue, colorlinks=true]{hyperref}
\usepackage{txfonts}
\usepackage{natbib}
\begin{document}

\title{A deep staring campaign in the $\sigma$ Orionis cluster}

   \subtitle{Variability in substellar members\thanks{Based on observations 
   made with ESO Telescopes at the Paranal Observatory under programme 
   ID 078.C-0042.}$^{,}$\thanks{Appendix~\ref{sec:appendix} is available 
   in electronic form at \url{http://www.aanda.org}.  The full 
   Table~\ref{tab:ukidss_vimos_cat} is available at the CDS via 
   anonymous ftp to \nolinkurl{http://cdsarc.u-strasbg.fr 
   (ftp://123.45.678.9)} or via 
   \url{http://cdsarc.u-strasbg.fr/viz-bin/qcat?J/A+A/XXX/XXX}.}}

  \author{P. Elliott\inst{1}
          \and
          A. Scholz\inst{2}
          \and
          R. Jayawardhana\inst{1}
          \and
          J. Eisl{\"o}ffel\inst{3}
          \and
         E. M. H\'ebrard\inst{1} 
          }
   \institute{Department of Physics and Astronomy, York University, Toronto, ON M3J 1P3, Canada
              \\              
              \email{pmelliott0@gmail.com}
   \and
   SUPA, School of Physics and Astronomy, University of St Andrews, North Haugh, St Andrews, Fife KY16 9SS, UK     
   \and
Th{\"u}ringer Landessternwarte Tautenburg, Sternwarte 5, D-07778 Tautenburg, Germany }

   \date{Received 18/04/2017; accepted 09/08/2017}

\abstract
{The young star cluster near $\sigma$\,Orionis is one of the primary environments to study the properties of young brown dwarfs down to masses comparable to those of giant planets.}
{Deep optical imaging is used to study time-domain properties of young brown dwarfs over typical rotational timescales and to search for new substellar and planetary-mass cluster members.}
{We used the Visible Multi Object Spectrograph (VIMOS) at the Very Large Telescope (VLT) to monitor a 24\arcmin $\times$ 16\arcmin field in the {\it I}-band. We stared at the same area over a total integration time of 21 hours, spanning three observing nights. Using the individual images from this run we investigated the photometric time series of nine substellar cluster members with masses from 10 to 60\,M$_\mathrm{Jup}$. The deep stacked image shows cluster members down to $\approx$5\,M$_\mathrm{Jup}$. We searched for new planetary-mass objects by combining our deep {\it I}-band photometry with public {\it J}-band magnitudes and by examining the nearby environment of known very low mass members for possible companions.}
{We find two brown dwarfs, with significantly variable, aperiodic light curves, both with masses around 50\,M$_\mathrm{Jup}$, one of which was previously unknown to be variable. The physical mechanism responsible for the observed variability is likely to be different for the two objects. The variability of the first object, a single-lined spectroscopic binary, is most likely linked to its accretion disc; the second may be caused by variable extinction by large grains. We find five new candidate members from the colour-magnitude diagram and three from a search for companions within 2000\,au. We rule all eight sources out as potential members based on non-stellar shape and/or infrared colours. The {\it I}-band photometry is made available as a public dataset.}
{We present two variable brown dwarfs. One is consistent with ongoing accretion, the other exhibits apparent transient variability without the presence of an accretion disc.  Our analysis confirms the existing census of substellar cluster members down to $\approx$7\,M$_\mathrm{Jup}$.  The zero result from our companion search agrees with the low occurrence rate of wide companions to brown dwarfs found in other works. }
 

   \maketitle
%

\section{Introduction}
\label{sec:intro}

The discovery of brown dwarfs in 1995 \citep{Nakajima1995, Rebolo1995}, in conjunction with the discovery of the first exoplanet around a solar-type star in the same year \citep{Mayor1995}, has triggered a significant revision in our ideas of star and planet formation. In particular, instead of a bimodal view where stars form from cores and planets in discs, our current picture is more complex, with brown dwarfs forming either `like stars' from the collapse of a core, helped by either turbulent fragmentation or dynamical encounters with more massive stars, or `like planets', that is, by disc fragmentation followed by ejection (see review by \citealt{Whitworth2007}). Hybrid scenarios where brown dwarfs form from gaseous clumps ejected from the disc may play a role as well \citep{Basu2012}.

Observational constraints for these theoretical developments have come from detailed and deep studies of nearby star forming regions (see review by \citealt{Luhman2012}). For each star, about 0.2-0.5 brown dwarfs are formed, in all regions studied so far, with only a minor, if any, dependence on environmental conditions  \citep{Scholz2013, Muzic2017}. Brown dwarfs can host massive discs \citep{Testi2016} and show signs of accretion, just like young stars. The widely-accepted view we have today is that most of the more massive brown dwarfs are an extension of the stellar mass function and form in a way similar to low-mass stars. 

The situation becomes much less clear for object masses approaching the planetary regime. The opacity limit for fragmentation at 5-10\,M$_\mathrm{Jup}$ is a principal barrier for star-like formation. Also, objects with Jupiter-like masses continue to grow through accretion, both in clouds \citep{Krumholz2016} and in wide orbits in discs \citep{Kratter2010}, which explains the paucity of free-floating objects with masses around or below the deuterium burning limit \citep{Scholz2012, Muzic2015}. Distinguishing between the various proposed scenarios is an important task for observers. The detailed characterisation of very young planetary-mass objects is challenging and still in progress. 

In this paper, we present deep optical imaging in the $\sigma$ Orionis cluster, obtained in a three night monitoring campaign with the visible multi-object spectrograph (VIMOS) at the very large telescope (VLT).  We use the sequence of deep images to look for variability in young brown dwarfs, following the previous work by, for example, \cite{Caballero2004}, \cite{Scholz2004}, \cite{Scholz2009}, \cite{Cody2010}, and \cite{Cody2011}. In contrast to these studies, our main focus is objects with estimated masses below or around the deuterium burning limit. By stacking our time series images we produced new, extremely deep optical images. We also used these deep images to search for new candidate members using available near-infrared photometry and by conducting a search for wide companions.

\section{Target region: The $\sigma$ Orionis cluster}

The young star cluster around the naked-eye star $\sigma$ Orionis 
harbours a rich population of young stars ranging from massive late O stars to very low-mass M dwarfs (e.g. \citealt{Garrison1967, Wolk1996, Walter1997, Sherry2004, Caballero2008}).  The compact size, negligble extinction and modest distance ($\sim 400$\,pc) make the $\sigma$ Orionis cluster an ideal ground to explore the initial mass function as well as the early evolution of young stellar and substellar objects. Deep surveys of the cluster revealed a rich population of brown dwarfs \citep{Bejar1999,Bejar2001} and free-floating objects with masses comparable to giant planets \citep{Barrado2001, Bihain2009, Zapatero2000}. The deepest large-area survey work in this cluster to date has been carried out based on multi-band photometry from the Visible and Infrared Survey Telescope for Astronomy (VISTA), the UKIRT Infrared Deep Sky Survey (UKIDSS) and other surveys (\citealt{PenaRamirez2012}; \citealt{Bejar2011, Lodieu2009}). Throughout this work we refer to, and cross match with, the young members and photometric candidates published in \cite{PenaRamirez2012}.  Their Tables~3, 5, and 7 give full details for all sources.

Previous distance estimates for the $\sigma$ Orionis cluster range from 300 to 450\,pc, which adds major uncertainties when inferring stellar parameters. The Hipparcos distance value is 352$^{+166}_{-168}$\,pc ($\pi = 2.84\pm 0.91$\,mas, \citealt{Perryman1997}). We used the parallaxes published in the Tycho-Gaia Astrometric Solution (TGAS) catalogue of the \textit{Gaia} DR1 \citep{GaiaDR1B2016, GaiaDR1A2016} to establish a new distance estimate for the $\sigma$ Orionis cluster. TGAS lists 24 stars within a 30\,\arcmin search radius of $\sigma$ Orionis; 15 of them have distances of 200-500\,pc based on TGAS, which makes them plausible cluster members. Eleven of those 15 appear in the Mayrit cluster member catalogue \citep{Caballero2008}. Their average parallax is $2.94\pm 0.40$\,mas, but excluding the faintest one (which has twice the average error) brings this to $2.84\pm 0.36$\,mas, corresponding to $352 \pm^{51}_{40}$\,pc. This is consistent with Hipparcos and encompasses most of the previous estimates.  It is also in line with recently published estimations using interferometric observations \citep{Schaefer2016} and TGAS data \citep{Caballero2017}. 

While the age of the $\sigma$ Orionis cluster is still somewhat uncertain, most authors agree that its population is significantly older than the Orion Nebula Cluster and somewhat younger than the nearest OB association Upper Scorpius, that is, between 1 and 10\,Myr. \cite{Hernandez2007}, \cite{Sherry2008}, and \cite{ZapateroOsorio2002} all arrive at age estimates of 2-4\,Myr, which are comparable to the ages adopted by the majority of the surveys mentioned above.  However, the updated age scale published by \cite{Bell2013} puts the cluster at 6\,Myr. In this work we use a distance of 352\,pc and an age of 5\,Myr when using the evolutionary models of \cite{Baraffe2003a} and \cite{Baraffe2015}.

\section{Observations and data reduction}
\label{sec:obs_data_reduc}

The data presented in this work were obtained on 24, 25, and 26 December 2006 using the VIMOS instrument at the VLT, Paranal.  The VIMOS instrument has four Charge-Coupled Devices (CCDs) each containing 2048 $\times$ 2440 pixels. The pixel scale is 0\arcsec.205, providing a field of view of $7\arcmin \times 8.3\arcmin$ in each quadrant. 

\begin{figure}
\begin{center}
\includegraphics[width=0.49\textwidth]{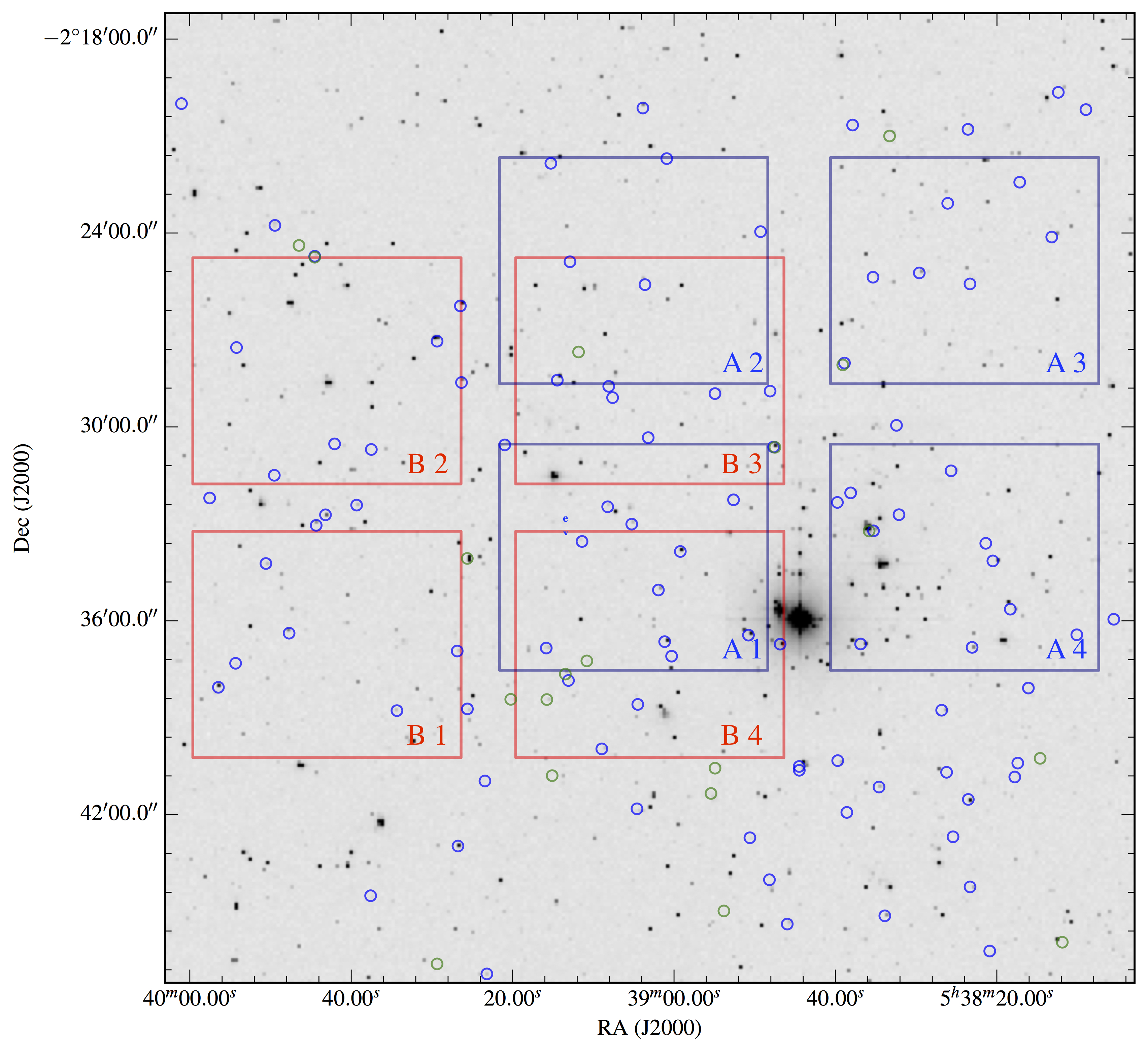}
\vspace{-0.4cm}
\caption{The observed pointings of our VIMOS/VLT observations overlaid on the 2MASS {\it J}-band image.  The navy blue and red rectangles are the four quadrants for Fields A and B, respectively.  The blue and green sources are young and photometric candidate sources, respectively, presented in \cite{PenaRamirez2012}. }
\label{fig:2mass_pointings}
\end{center}
\end{figure}

The observations spanned approximately seven hours each night, and alternated between the two pointings, presented in this work as Fields A and B. Figure~\ref{fig:2mass_pointings} shows the observed pointings overlaid on the Two Micron All Sky Survey (2MASS) {\it J}-band image. Each individual exposure was 300\,s long and taken in the {\it I}-band.  The final dataset was comprised of 112 science images for each of the pointings and additionally 15 dark sky exposures from the three nights.   The reduction of the data was done for each of the separate quadrants to account for any CCD-dependent behaviours.

{\begin{table*}
\begin{center}
\tiny
\caption{Basic properties of young members and photometric candidates in $\sigma$ Orionis from \cite{PenaRamirez2012} that are studied in this work.}
\begin{tabular}{p{3.5cm} p{1.1cm} p{1.1cm} p{1.cm} p{1.9cm} p{0.8cm} p{1.1cm} p{0.7cm}  p{2.9cm}}
\hline\hline\\[-2ex]
  \multicolumn{1}{l}{Resolvable Simbad ID} &             
  \multicolumn{1}{l}{RA} &
    \multicolumn{1}{l}{DEC} &
        \multicolumn{1}{l}{Feat.~\tablefootmark{a}} &
        \multicolumn{1}{l}{IR Excess} &
        \multicolumn{1}{l}{{\it J~\tablefootmark{b}}} &        
        \multicolumn{1}{l}{{$\mathcal{M}$}} &                
        \multicolumn{1}{l}{{Var?~\tablefootmark{c}}} &            
         \multicolumn{1}{l}{Comments} \\         
  \multicolumn{1}{l}{} &             
  \multicolumn{1}{l}{hh:mm:ss.s} &    
  \multicolumn{1}{l}{dd:mm:ss} &
  \multicolumn{1}{l}{} &
  \multicolumn{1}{l}{} &
  \multicolumn{1}{l}{(mag)} &  
  \multicolumn{1}{l}{(M$_\mathrm{Jup}$)} &  
  \multicolumn{1}{l}{} &    
  \multicolumn{1}{l}{} \\      
  \hline\\[-1.6ex]
  \multicolumn{9}{c}{Variability and deep image analysis} \\        
  \hline\\[-1.6ex]
Mayrit 258337 & 05:38:38.1 & -02:32:03 & RV, g, d &     Y(4.5, 8.0, 12.0) & 15.07  & 56 & Y  & SB1, Known var.\\
Mayrit 396273  & 05:38:18.3 & -02:35:39 & RV, g & N & 15.29 & 47 & Y\\
Mayrit 379292  & 05:38:21.4 & -02:33:36 & RV, Li, g, d &Y(12.0)&15.31 & 47&  \ldots  \\
{[MJO2008]} J053852.6-023215  & 05:38:52.6 & -02:32:15 & RV, g& N & 16.18 & 29 &  \ldots  \\
{[BNM2013]} 90.02 782  & 05:39:12.9 & -02:24:54 & H$\alpha$ & N & 16.68 &24&  \ldots  \\
{[BNM2013]} 90.02 1834  & 05:39:00.3 & -02:37:06 & H$\alpha$, d & Y(4.5, 8.0) &17.19 & 19 &  \ldots  \\
{[BZR99]} S Ori 51   & 05:39:03.2 & -02:30:20 & g& N & 17.16 & 19 &  \ldots & \\
{[BZR99]} S Ori 50  & 05:39:10.8 & -02:37:15 & \ldots & N &  17.47 & 17 &  \ldots & Photometric cand.  \\
{[BZR99]} S Ori 58  & 05:39:03.6 & -02:25:36 & H$\alpha$, d &Y(4.5, 8.0)& 18.42 & 11 &  \ldots \\
  \hline\\[-1.6ex]
  \multicolumn{9}{c}{Deep image analysis only} \\        
  \hline\\[-1.6ex]
{[BZR99]} S Ori 60 & 05:39:37.5  &  -02:30:42  & H$\alpha$, d &         Y(8.0) & 19.02  &  8 & \ldots & \\
{[BZR99]} S Ori 62  & 05:39:42.1  &  -02:30:32  &    H$\alpha$ &  N       & 19.14  &  8 & \ldots  & \\  
{[BZR99]} S Ori 65 & 05:38:26.1  &  -02:23:05  & d &    Y(4.5, 8.0) & 20.30  &  5 & \ldots  & \\
\hline\\[-0.4ex]
\end{tabular}
\tablefoot{\tablefoottext{a}{RV: Radial velocity consistent with systemic cluster velocity, g: low-gravity atmosphere, Li: Lithium absorption, H$\alpha$: Strong, broad H$\alpha$ emission, d: Presence of a disc.}\tablefoottext{b}{{\it J}-band magnitude from \cite{PenaRamirez2012}. }
\tablefoottext{c}{Indicates if variability was identified from the analysis in this work.}}
\label{tab:young_sources}
\end{center}
\end{table*}}

The observing conditions were variable throughout the three nights of observations.  The median seeing and standard deviation of the science frames from each night was 0.84$\pm$0.20, 0.89$\pm$0.66, and 1\arcsec.23$\pm$0.66. This large variation in seeing limited our ability to remove the small-scale fringe structure resulting from night sky emission. We adopted the same approach that is used in \cite{Alcala2002}, \cite{LopezMarti2004},  and \cite{Scholz2005a} for the data reduction.

In Table~\ref{tab:young_sources} we list all of the young members and photometric candidates from \cite{PenaRamirez2012} that are covered in the analysis presented in this work.

\section{Photometry}
\label{sec:phot_ast}

\subsection{Time series of relative photometry}
\label{subsec:time_series_phot}

To extract the sources' photometry and astrometry from our images, we used the astropy-affiliated photutils  package \citep{photutils} in Python. Many of the extraction algorithms used within this package are the same as those used by SExtractor \citep{Bertin1996}.  

The first step was to build an input catalogue of sources for each quadrant of each pointing. We selected the image with the best seeing and extracted sources using a 5\,$\sigma$ criterion.  We then used this list of positions to extract the astrometry and photometry of sources in the time series of images.  To account for any potential pixel drift between individual observations, we re-centred our apertures before extracting the photometry in each individual exposure.  The typical pixel drifts were $<$ 2 pixel for all our observations.  In order to account for variation in the background signal, we performed a local background subtraction for each extracted source.  We calculated the average flux within an eight-pixel-width annulus around each source. The inner edge of this annulus was set by two times the Full-Width Half Maximum (FWHM) of each individual observation.  We multiplied this average flux value by the area of the source's aperture and subtracted the resultant value.  As our input catalogue was constructed from the image with the best seeing, the value of flux for very faint sources in exposures with much worse seeing could be 0 or below, due to local background subtraction.  In such cases the photometry for these apertures was not included in our final analysis. 

We removed any images that had a seeing (FWHM) at the time of observation $>$1\arcsec.5 (23 exposures).  This was to avoid neighbouring sources contaminating the background flux calculated in the annulus around each source.  
The analysis presented in this section was first performed using all exposures. However, the precision in magnitude was severely limited and, therefore, the images with the worst seeing were rejected, leaving 79\% of the original dataset.\\

Primarily, we were interested in differential photometry in our analysis, therefore absolute calibration of magnitudes was not necessary. Our approach was to use a sample of reference stars in each quadrant of each pointing, initially selected by their low standard deviation, to account for any variations in magnitude caused by the changing observing conditions.  The process was as follows.

\begin{figure}
\begin{center}
\includegraphics[width=0.49\textwidth]{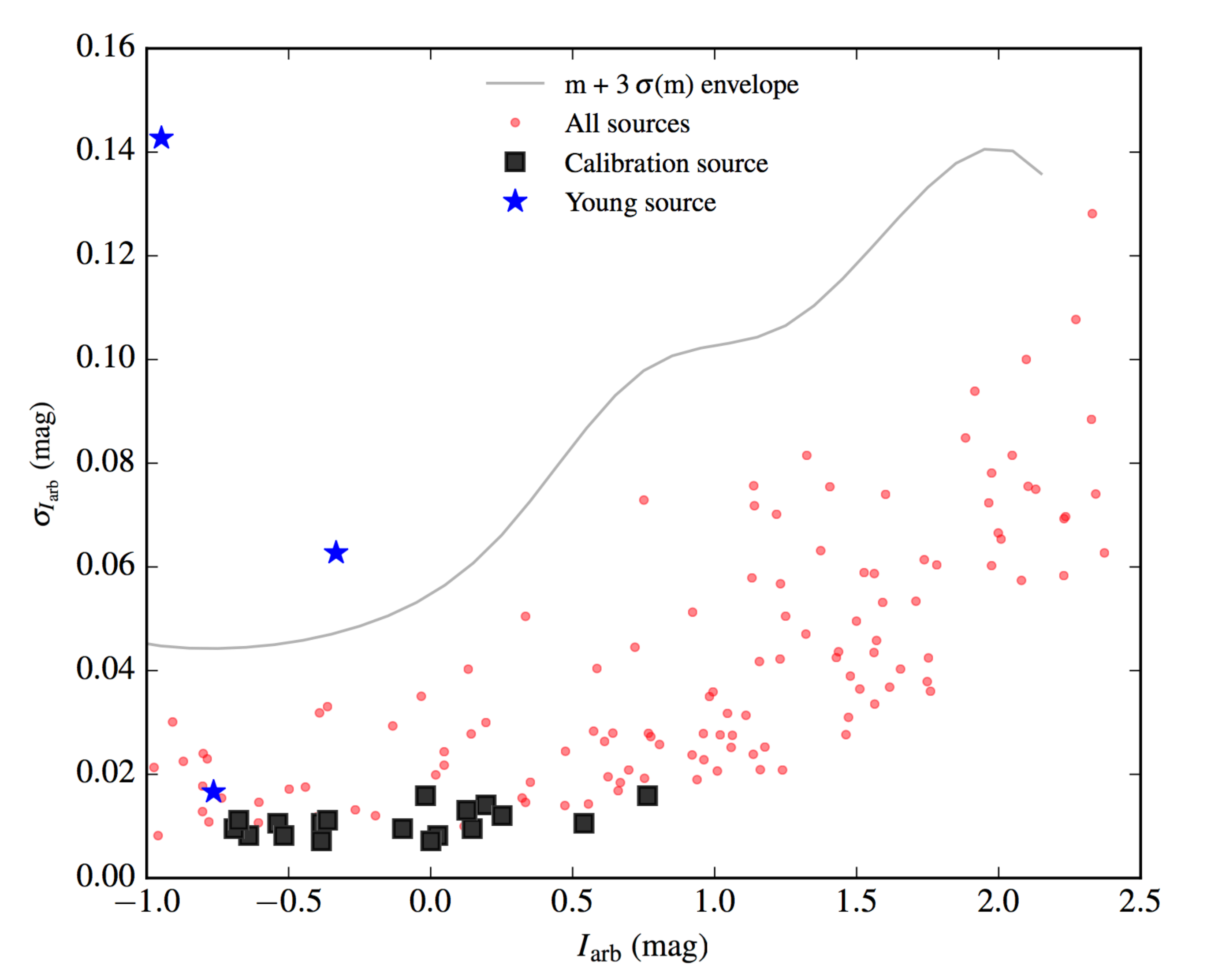}
\vspace{-0.7cm}
\caption{Standard deviation of the light curves in Field A, quadrant 1 for $\sigma$ Orionis versus median uncalibrated {\it I}-band magnitude.} 
\label{fig:rms_mag}
\end{center}
\end{figure}

We first selected bright, unsaturated sources with a low standard deviation compared to the other sources in the same pointing.  We plotted the light curves of all of these sources and removed any that were obviously inconsistent with the others, that is sources with significant changes in brightness not seen in other sources.  We then calculated a median light curve from the remaining sources, subtracted this median from each of the individual light curves, excluded 3 $\sigma$ outliers, and calculated the sum of the residuals for each.  If the residuals of the light curve were $<$0.03\,mag, the light curve remained in the reference stars list; if not, it was omitted and the procedure was run again using the new list as an input.  The criterion of 0.03\,mag was a compromise between having enough reference stars (typically ten or more) and achieving a high precision.  The mean precision of our master reference light curves for all pointings is between 11 and 19 mmag, comparable, albeit worse, than the 7 mmag achieved in \cite{Scholz2005a}. These master light curves were finally subtracted from the light curves of all individual sources in the pointing.  Figure~\ref{fig:rms_mag} shows our results for Field A, quadrant 1.  

\subsection{Stacking individual images}

The second aspect of the analysis presented in this work is the stacking of individual exposures to create a deep image of the $\sigma$ Orionis region in $I$ band. In order to account for the small variations (typically $<$2 pixel) of source positions between observations from the three nights, we applied small shifts relative to the first observation frame. The shifts were calculated by extracting the positions of all the sources in each image and subtracting these from the counterparts in the first observation frame.  A median of all the resultant residuals was taken and this was subtracted from each respective image.  The final deep image for each pointing was a median taken from the realigned individual exposures which had a seeing value $\leq$1.5\arcsec.  The standard deviation in the background signal for the resultant set of deep images was approximately 0.5\% of the average background compared to $\approx$2\% for individual images.  This results in the ability to detect sources $\approx$1.5\,mag fainter in our deep image.  This is highlighted by the three very low-mass sources (lowest-mass source $\approx$5\,M$_\mathrm{Jup}$) in Table~\ref{tab:young_sources} that we recover.  We applied an offset of 31\,mag, an arbitrary choice, to produce small $I$ - $J$ values, which gives us a 5\,$\sigma$ limit of 22.5\,mag for our deep image. Our uncalibrated I-band photometry for all sources (2139) successfully cross-matched with UKIDSS/DR9 GCS catalogue \citep{Warren2007} is available publicly via the VizieR service.

\section{Variability of sources}
\label{sec:variability}

\subsection{Detection of variability in our observations}

We searched for variability among extracted sources using the standard deviation of the lightcurves, shown in Figure~\ref{fig:rms_mag}. In short, we identified sources with standard deviation significantly larger than the noise in similarly bright sources. First we calculated the median value ($m$) and standard deviation ($\sigma_\mathrm{e}(m)$) of the values as a function of magnitude.

We did this in 0.5 magnitude bins, ensuring there was a sufficient number of data points (typically $>$10--20) in each bin to calculate the relevant statistics. We then calculated an upper envelope ($e$) for each magnitude bin consisting of the median value plus three times the standard deviation in each bin. We performed cubic interpolation to create a finer grid of $m$, $\sigma_\mathrm{e}(m)$ values. An example of this upper envelope is shown as the grey line in Figure~\ref{fig:rms_mag}.  With this measurement we were able to quantitatively assess whether an individual source's standard deviation ($\sigma_\mathrm{m}$) was significant with respect to other sources of approximately the same magnitude. Equation~\ref{eq:sig_crit} shows our derived variability quantity ($\Lambda$).  We define a source as variable if $\Lambda > 1$: 

\begin{equation}
\label{eq:sig_crit}
\Lambda = \frac{m + 3\,\sigma_\mathrm{e}(m)}{\sigma_\mathrm{m.}}
\end{equation}

{
\begin{table}
\tiny
\caption{Photometric properties of all variable sources identified in this work as well as all non-variable properties from either young or photometric candidate members of $\sigma$ Orionis from \cite{PenaRamirez2012}.}
\begin{tabular}{p{3.5cm}p{0.9cm} p{0.9cm} p{1.1cm}  p{0.6cm}}
\hline  \hline\\[-1.6ex]
  \multicolumn{1}{l}{ID} &
  \multicolumn{1}{l}{Arb. {\it I}} &
  \multicolumn{1}{l}{$\sigma_\mathrm{m}$\tablefootmark{a}} &  
  \multicolumn{1}{l}{$m + 3\sigma_\mathrm{e}$\tablefootmark{a}} &    
  \multicolumn{1}{l}{$\Lambda$\tablefootmark{a}} \\
  \multicolumn{1}{l}{} &    
  \multicolumn{1}{l}{(mag)} &
  \multicolumn{1}{l}{(mag)} &     
  \multicolumn{1}{l}{(mag)} &       
  \multicolumn{1}{l}{} \\
  \hline\\[-1.6ex]
  \multicolumn{5}{c}{All variable sources} \\
  \hline\\[-1.6ex]
Mayrit 258337 & -0.95 & 0.143 & 0.051 & 2.804   \\
Mayrit 396273 & -0.33 & 0.063& 0.049 & 1.291  \\
UKIDSS 442414579467\tablefootmark{b} & 0.09 & 0.170 & 0.130 & 1.308  \\
UKIDSS 442414579362\tablefootmark{c} & -0.11 & 0.033 & 0.028 & 1.193 \\
  \hline\\[-1.6ex]
  \multicolumn{5}{c}{Non-variable from Y and C catalogues} \\
  \hline\\[-1.6ex]
Mayrit 379292 & -0.76 & 0.017 & 0.059  & 0.280  \\
{[MJO2008]} J053852.6-023215 & 0.36 & 0.024& 0.138  & 0.177  \\
{[BNM2013]} 90.02 782 & 1.20 & 0.025 & 0.095 & 0.266 \\
{[BNM2013]} 90.02 1834 & 1.55 & 0.174  & 0.221 & 0.787  \\
{[BZR99]} S Ori 51 & 1.94 & 0.053 & 0.197 & 0.272  \\
{[BZR99]} S Ori 50 & 2.13 & 0.085 & 0.268 & 0.316  \\
{[BZR99]} S Ori 58 & 3.62 & 0.119  & 0.508 & 0.129  \\
\hline
\end{tabular}
\tablefoot{\tablefoottext{a}{Properties described in Equation~\ref{eq:sig_crit}.}  \tablefoottext{b}{Background field star: 05:39:04.2, -02:31:11} \tablefoottext{c}{Background field star:  05:39:07.8, -02:30:55}.}
\label{tab:rms_young_sources}
\end{table}}

The calculated values are shown in Table~\ref{tab:rms_young_sources}.   In the case that a single source had time series photometry from two separate pointings, we visually checked the two time series to identify any potential discrepancies. We proceeded with one of the two in further analysis given there were no significant differences.
From the analysis of all pointings, four sources were initially classed as variable, two of which have been previously classified as young members of $\sigma$ Orionis, both shown in Figure~\ref{fig:rms_mag}. %

\begin{figure}
\begin{center}
\includegraphics[width=0.49\textwidth]{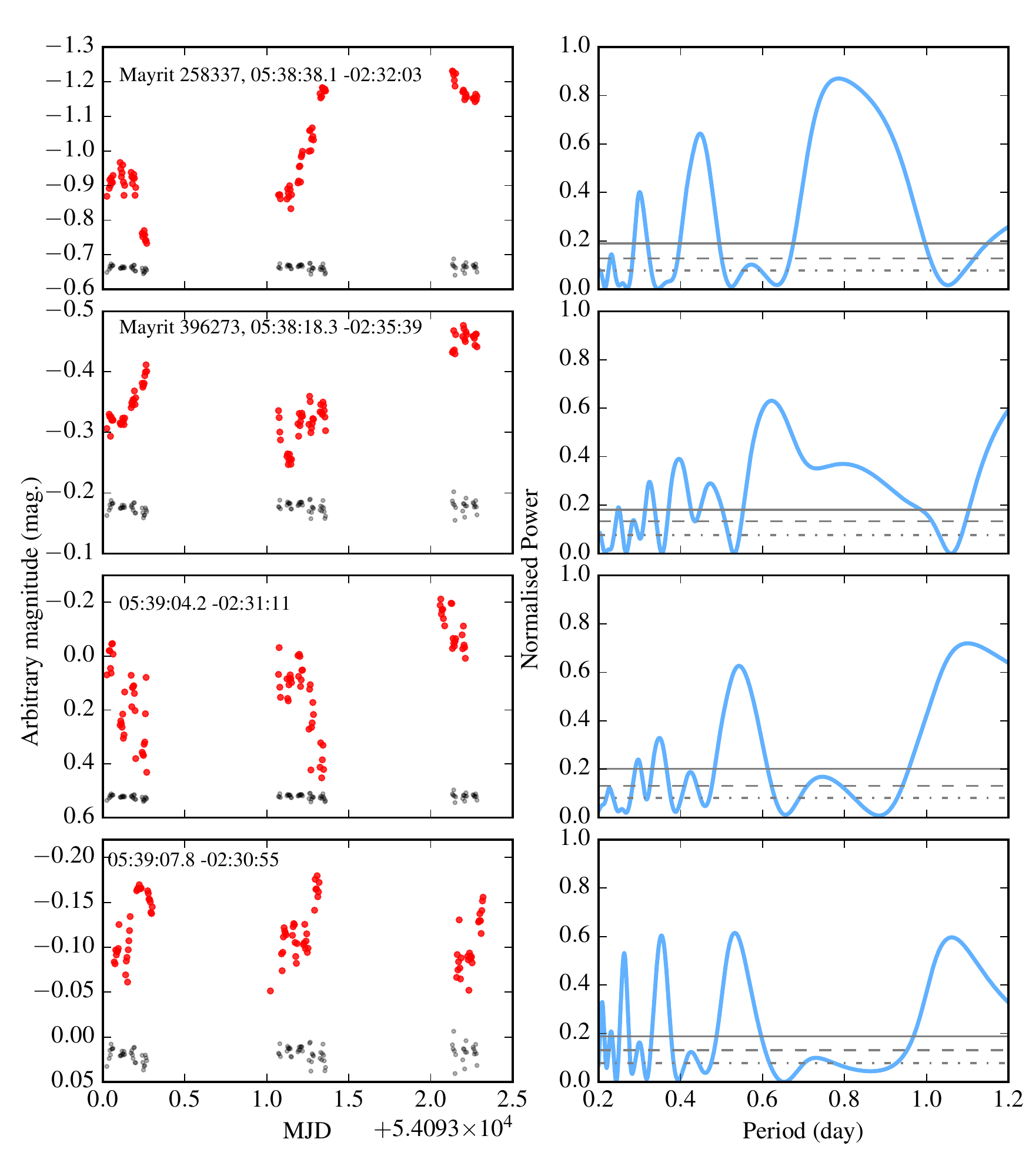}
\vspace{-0.3cm}
\caption{Left panels: VIMOS light curves for the four variable sources identified in this work (Table~\ref{tab:rms_young_sources}).  The red markers are the light curves of each source, the grey markers are the light curve of a calibration source of similar magnitude, shown for comparison. Right panels: Generalised Lomb Scargle periodograms for each variable source.  The dot-dashed, dashed, and solid grey lines are 1\,$\sigma$, 2\,$\sigma$, and 3\,$\sigma$ intervals from 1000 bootstrap samples.}
\label{fig:variable_light_curves}
\end{center}
\end{figure}

\subsection{Bona-fide young sources}

In this section we discuss our {\it I}-band observations, AllWISE mid-infrared data, and any noteworthy properties from previous studies to summarise the variable properties of the two young, variable objects more fully.
Young stellar objects often show photometric variability due to a variety of effects, ranging from stellar activity (spots, flares), accretion, variable extinction along the line of sight, and obscurations by dust features in the disc.  The variability can range from strictly periodic to irregular  \citep{Cody2014}. The dominant timescale in the variations is usually the rotation period. For brown dwarfs at the age of $\sigma$ Orionis, typical periods are in the range of 1-3\,d \citep{Scholz2015}, comparable to the duration of our observations. 

One of the two variable objects with evidence of youth (see first two rows in Table\,\ref{tab:rms_young_sources}) is a newly discovered variable brown dwarf. The inter-night variations in our data are significantly larger than the variability within one night, indicating that typical timescales of the observed variations are indeed in the range of days, comparable to rotational cycles.  The relative magnitudes in both objects do not show any trend with airmass, thus, the variability is most likely intrinsic to the objects and not related to atmospheric extinction.  The light curves and Generalised Lomb-Scargle periodograms (GLS; \citealt{Zechmeister2009}) of the two young variable sources and the two older sources are shown in the left and right hand panels of Figure~\ref{fig:variable_light_curves}, respectively. 

The left panels of Figure~\ref{fig:wise_light_curve} show the AllWISE ({\it W1}: 3.4\,$\mu$m, {\it W2}: 4.6\,$\mu$m) data taken from the Multiepoch Photometry Table\footnote{\url{http://irsa.ipac.caltech.edu/cgi-bin/Gator/nph-scan?mission=irsa&submit=Select&projshort=WISE}.} for the two young objects. The data were taken at two different epochs covering the approximate date ranges 9-11 March 2010 and 16-18 September 2010.  In the first of the right panels we show the results of the Pearson correlation coefficient (correlating {\it W1} and {\it W2}) for each object.  Data points without measurement uncertainties or with low signal to noise ($<$5) have been removed.  To take the measurement uncertainties into account, we created 1000 synthetic arrays from random realisations of a Gaussian distribution centred on each value and using its respective measurement uncertainty as the width of the distribution.  The histograms show the results from these 1000 samples for both the $r$ and $p$ value for each object.  The $r$ value is a measure of the correlation between the two magnitude arrays, and the $p$ value a statistic of how likely it is that uncorrelated data could produce such an $r$ value.  The most right hand panels of Figure~\ref{fig:wise_light_curve} show the GLS periodograms for both sets of {\it W1} and {\it W2} data. 

\subsubsection{Mayrit 258337}

The object Mayrit 258337 was first classified as variable by \cite{Lodieu2009} based on two epochs of photometry. \cite{Hernandez2007} reported mid-infrared excess and thus clear evidence for the presence of a disc (no. 633 in their catalogue).  This object is also a known single-lined spectroscopic binary \citep{Maxted2008}.   With a system mass of about 56\,M$_\mathrm{Jup}$, estimated from unresolved photometry, the masses of the individual components are lower than that. The primary is most likely $\sim$40\,M$_\mathrm{Jup}$ and the secondary $\sim$25\,M$_\mathrm{Jup}$, assuming a 0.7\,mag difference.

The light curve peak-to-peak amplitude is $\approx$0.5\,mag for Mayrit 258337 (see Figure~\ref{fig:variable_light_curves}).  From the GLS periodogram analysis of these {\it I}-band observations, we found a series of significant peaks.  The peak with the highest power is at 0.77\,day.  We tried to use this period to fit the data. However, there was clearly still a lot of higher-order features in the variability signature.  Therefore, at this time, it is unclear whether this peak at $\sim$0.77\,day is related to the rotation of one of the objects in the system.

\begin{figure*}
\begin{center}
\includegraphics[width=0.98\textwidth]{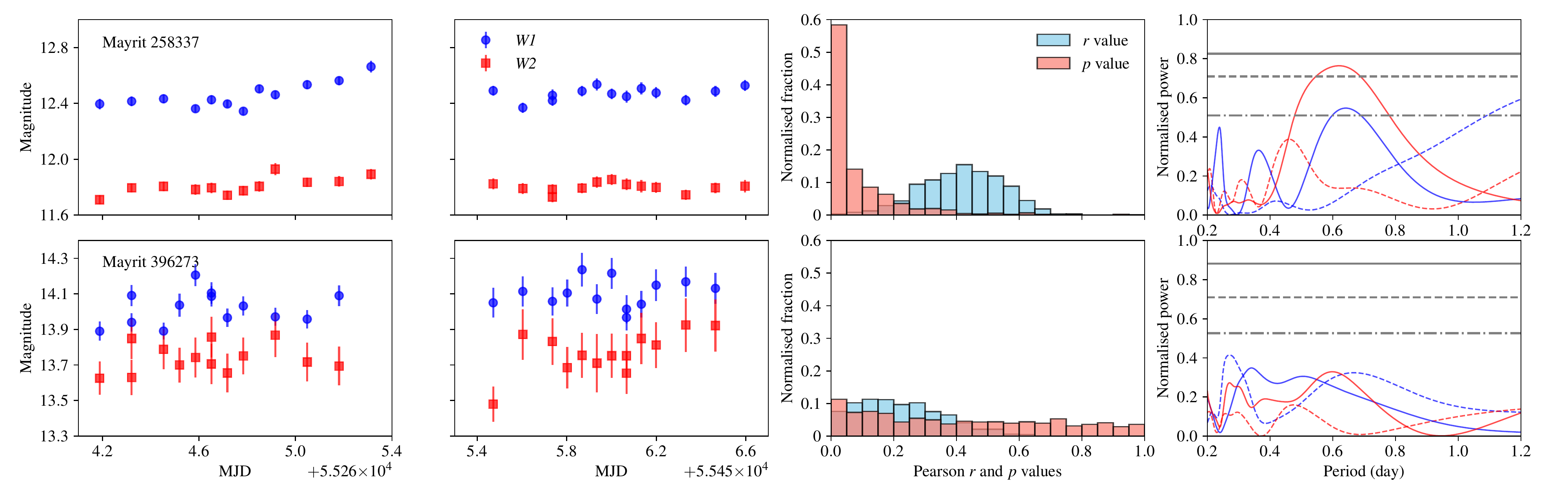}
\vspace{-0.2cm}
\caption{Left panels: Photometry from the AllWISE Multiepoch Photometry Table for Mayrit 258337 (upper panels) and Mayrit 396273 (lower panels), separated by MJD range.  The blue circles and red squares represent the photometric bands {\it W1} and {\it W2}, respectively.  First right panels: Pearson correlation coefficients ($r$) and their respective $p$ values for 1000 simulations sets of \textit{WISE} magnitudes. Second right panels: Generalised Lomb-Scargle periodograms for each MJD versus \textit{WISE} magnitude array. Colours are the same as in left panels.  Dotted and solid lines present the first and second MJD range, respectively. The 1\, 2, and 3\,$\sigma$ intervals from 1000 bootstrap samples (for {\it W2}) are shown as the grey dot-dashed, dashed, and solid lines, respectively.}
\label{fig:wise_light_curve}
\end{center}
\end{figure*}

In the first of the right hand panels of Figure~\ref{fig:wise_light_curve}, we can see that there is significant correlation between the {\it W1} and {\it W2} magnitudes for this object.  This is shown by the consistently low ($\lesssim$0.05) $p$ values in the synthetic samples and average $r$ value of $\approx$0.5.  A GLS periodogram of the two series of \textit{WISE} magnitudes (second right hand panel of Figure~\ref{fig:wise_light_curve}) shows that we found one marginally significant period ($\approx$2.5$\sigma$, $\approx$1\% false alarm probability) at 0.63\,day in the {\it W2} magnitude for the second Modified Julian Date (MJD) range.  Additionally, a similar period is found (at a much higher false alarm probability of $\approx$30\%) in the {\it W1} in the same MJD range. Given the short span of these observations, it is hard to conclude definitely on this recovered periodic signal.  However, it may be that the 0.77\,day signal from our observations and the 0.63\,day signal from AllWISE data are related to the rotation of the object. Given the width of the peaks ($\pm$0.2\,day) in the periodograms, the values are consistent.

\subsubsection{Mayrit 396273}

Mayrit 396273, first reported in \cite{Bejar2004}, has no mid-infrared excess out to 8$\,\mu m$ (no. 446 in the list by \citealt{Hernandez2014}). 
The AllWISE \citep{Wright2010, Mainzer2011} database contains fluxes at 12 and 22$\,\mu m$ for this source, with a signal-to-noise ratio of 14 and 4, respectively, but the detections do not look convincing in the images and are likely contaminated by the source's neighbours. This could mean that the object is either disc-less or has a depleted disc or a disc with an inner hole. The object does have an X-ray detection, reported by \cite{Franciosini2006} and \cite{Caballero2010b},  indicating strong magnetic activity. A light curve for this object was obtained by \cite{Cody2010}, but no variability was found, although their precision was comparable or better than ours. Thus, the variability appears to be transient.  

The light curve peak-to-peak amplitude is $\approx$0.25\,mag for Mayrit 396273 in our {\it I}-band observations.  It exhibits a shallow dip at the beginning of the second night, maybe related to an eclipse. The GLS periodogram of these data has a number of significant peaks; the peak with the highest power is at $\approx$0.61\,day.  However, as in the case of Mayrit 258337, using this peak in a sinusoidal fit did not describe the data well and therefore should be treated with caution.

In Figure~\ref{fig:wise_light_curve} we see that the average $r$ value is lower ($\approx$0.2) and poorly constrained, as shown by the wide and uniform-like distribution in the $p$ value.  In other words, uncorrelated {\it W1} and {\it W2} data could produce such an $r$ value in a significant number of simulated samples.  Additionally, no significant periodicity was found for Mayrit 396273 (see the furthest right panel of Figure~\ref{fig:wise_light_curve}).

\subsection{Young brown dwarfs with strong irregular variability}
\label{sec:variable_brown_dwarfs}

Variability due to magnetic spots is usually periodic or quasi-periodic over at least a few rotational cycles, which is not the case for our two variable young sources. The typical shape of flares is
also not seen in our light curves. Thus, magnetic activity can be excluded as the sole origin for the variations.

\begin{figure}
\begin{center}
\includegraphics[width=0.49\textwidth]{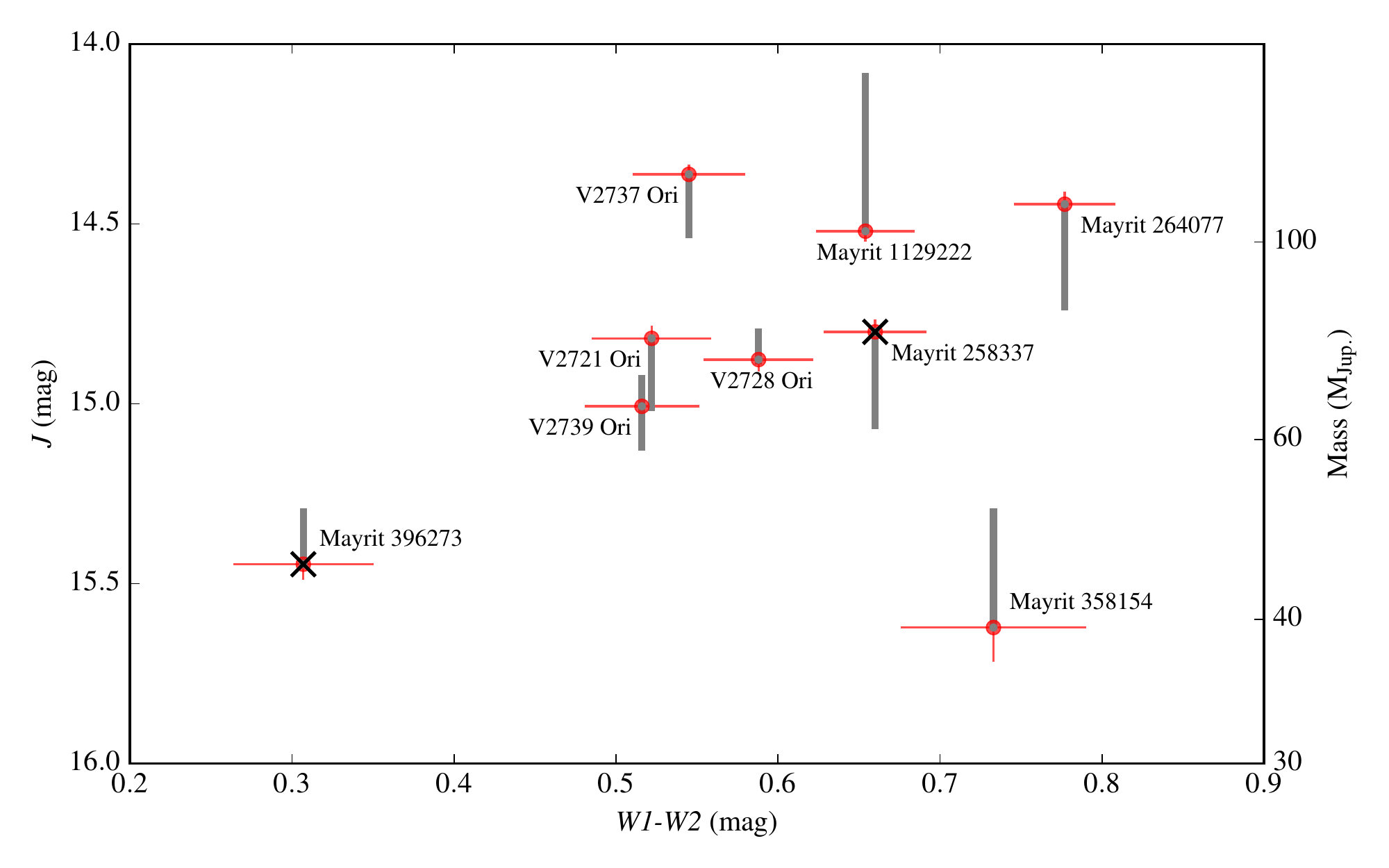}
\vspace{-0.2cm}
\caption{{\it W1-W2} versus 2MASS {\it J}-band magnitude for low-mass variable targets listed in Table~\ref{tab:young_variable}.  The thick  grey lines represent the largest difference between 2MASS {\it J} values and those of \cite{PenaRamirez2012} and UKIDSS.  The red lines are the 2MASS and AllWISE measurement uncertainties.  The black crosses indicate the two variable objects identified in this work.}
\label{fig:wise_jmag}
\end{center}
\end{figure}

To constrain the nature of our two variable young brown dwarfs, we compare them with similar, previously known objects. Since the discovery of the $\sigma$ Orionis cluster, the region has been targeted multiple times in deep photometric monitoring campaigns. We have prepared a list of very low-mass objects with strong and partially irregular variability in the $\sigma$ Orionis cluster. In Table~\ref{tab:young_variable} we include all known objects in this cluster with evidence of youth, with masses below or around the sub-stellar limit ($J\geq14$ mag in the $I$ band or spectral type mid M or later), and with strong variability caused by cool spots co-rotating with the objects due to magnetic activity. All given amplitudes are measured peak-to-peak and have been observed in the {\it I} band, if not otherwise indicated.  Figure~\ref{fig:wise_jmag} shows the maximum difference in \textit{J}-band magnitude from three catalogues as a function of {W1-W2} colour for all sources in Table~\ref{tab:young_variable}.  The range of \textit{J}-band magnitudes should be treated as lower envelopes.  Typical uncertainties for objects with $J$ magnitudes in the range 14 - 16 mag are $\approx$0.01\,mag for both UKIDSS and 2MASS.

{
\begin{table*}
\caption{Census of low-mass (spectral types mid M or later) members of the $\sigma$ Orionis cluster that show strong and irregular variability.}
\begin{tabular}{p{2.2cm} p{0.8cm}  p{2.8cm} p{0.8cm} p{0.6cm} p{0.8cm}  p{3.cm}p{3cm}}
\hline  \hline\\[-1.6ex]
  \multicolumn{1}{l}{Name} &
    \multicolumn{1}{l}{SpT} &
  \multicolumn{1}{l}{{\it J~\tablefootmark{a}}} &
  \multicolumn{1}{l}{{\it {W1-W2}}} &  
  \multicolumn{1}{l}{acc.} &  
  \multicolumn{1}{l}{IR excess} &    
  \multicolumn{1}{l}{Amplitude range} &
    \multicolumn{1}{l}{References} \\
  \multicolumn{1}{l}{} &    
  \multicolumn{1}{l}{} &
  \multicolumn{1}{l}{(mag)} &  
    \multicolumn{1}{l}{(mag)} &    
  \multicolumn{1}{l}{}  &
  \multicolumn{1}{l}{(mag)} &
  \multicolumn{1}{l}{(mag)} &     
  \multicolumn{1}{l}{} \\
  \hline\\
V2737 Ori & M4   &14.46 / 14.54 / 14.36   & 0.55 &yes & yes    &  0.7-1.2    & a (SE2004 33), b, c, d\\
V2721 Ori  &M4   & 14.92 / 15.02 / 14.82  & 0.52 &yes  &yes     & 0.2-0.7       & a (SE2004 2), b, c\\
V2739 Ori  &M5   & 14.92 / 15.13 / 15.01 & 0.52 & yes  & yes     & 0.4-0.5 & a (SE2004 43), c\\
V2728 Ori  & M6  & 14.89 / 14.79 / 14.88 & 0.59 & yes  & yes     & 0.2-0.7        & d, e, f, g \\
Mayrit 1129222   & \ldots &  14.18 / 14.08 / 14.52 &0.65 & ?  &  yes &    2.0 & d, h \\
Mayrit 358154  & M5 & 15.29 / 15.34 / 15.62 & 0.73 &yes& yes &    0.9  & d, h \\
Mayrit 264077 &  M3 & 14.74 / 14.46 / 14.45 & 0.78 &yes &yes  &   0.9       &  h  \\

  \hline\\[-1.6ex]
Mayrit 258337     & \ldots  &  15.07 / 14.81 / 14.80 & 0.66  & yes & yes   &   0.5                 &         d, i \\
Mayrit 396273    &  \ldots &  15.29 / 15.29 / 15.45   & 0.31   & ? & no    &  0.25      &                    i \\
\hline\\[-0.1ex]
\end{tabular}
\tablefoot{\tablefoottext{a}{Three magnitude values given in order of \cite{PenaRamirez2012}, UKIDSS, and 2MASS \citep{Cutri2003}.}}
\tablebib{{a: \cite{Scholz2004}, b: \cite{Scholz2009b}, c: \cite{Bozhinova2016}, d: \cite{Lodieu2009}, e: \cite{Caballero2006}, f: \cite{Caballero2004}, g: \cite{Cody2010}, h: \cite{Cody2010}, i: this work.}}
\label{tab:young_variable}
\end{table*}}

Three of these sources were originally found by \cite{Scholz2004}. From \cite{Cody2010}, we selected three objects with amplitudes $>0.2$\,mag.  We note that their catalogue contains a number of other objects with low-level irregular variations. Additionally, \cite{Scholz2004} identified a few more faint sources that would satisfy the criteria for Table~\ref{tab:young_variable} (SE16, 85, 95).  Their variations, however, are close to the photometric noise and are so far not confirmed. We exclude these three from further considerations but note they require further attention. Additional low-level variables with possibly irregular contributions were published by \cite{Caballero2004} and \cite{Bailer-Jones2001}.

In Figure~\ref{fig:wise_jmag} we plot a near and mid-infrared colour magnitude diagram of the sample in Table~\ref{tab:young_variable}. On the y-axis, the {\it J}-band magnitude is a proxy for stellar luminosity and, thus, mass. For these highly variable objects, these stellar parameters are highly uncertain and without understanding the origin of the variability it is difficult to distinguish between low-mass stars and brown dwarfs. On the x-axis, the $W1-W2$ colour shows possible colour excess from circumstellar or substellar dust, which is evidence of a disc.

This diagram, in addition to the information in Table~\ref{tab:young_variable}, puts our two variables 
in context. Mayrit 258337 has mid-infrared color excess similar to the other known irregular variables in $\sigma$ Orionis, persistent variability over multiple years, and also a similar photometric amplitude over the timescale of days.  
Mayrit 396273, however, is an outlier in Figure~\ref{fig:wise_jmag} as it does not show significant infrared excess.

The correlation between {\it W1} and {\it W2} and the periodic nature of the signal for Mayrit 258337 matches the variability characteristics observed in many prototypical T Tauri stars \citep[see][]{Bertout1989, Rigon2017}. Thus, the variability is most likely caused by accretion or processes in the inner disc. Mayrit 258337 is one of the lowest-mass objects known that falls into this well-studied T Tauri category.

As already mentioned, Mayrit 396273 is the exception in our sample.  Its strong variability is transient, which also sets it apart from most of the other sources. Additionally, there is no significant correlation between its {\it W1} and {\it W2} magnitudes, nor any significant periodicity as a function of time, and its relative measurement uncertainties are large.  The object joins a group of recently identified young (5-10\,Myr) very low-mass objects without primordial accretion discs that still show strong variability on timescales of hours and days. \cite{Bozhinova2016} presented spectrophotometry for four young mid-M stars in the nearby $\epsilon$\,Ori region, without evidence of accretion; two of them without infrared excess. These objects, originally discovered by \cite{Scholz2005}, show little spectral variations, although the flux changes by 0.1-0.6\,mag in the $I$ band over the course of four observing nights.  \cite{Bozhinova2016} concluded that their variations may be caused by variable extinction by large grains, located in an evolved, clumpy disc. Other objects with similar basic characteristics  have recently been found in the Upper Scorpius star forming region based on light curves from K2 \citep{David2017, Stauffer2017}. These authors suggest several scenarios to explain the variability, including obscurations by dusty debris, by warm coronal gas clouds, or by clouds associated with a close-in planet. A detailed comparison of this family of variable objects is beyond the scope of this study and reserved for a future paper.  Mayrit 396273 may be the lowest mass counterpart of this type of object found so far, with an estimated mass well in the substellar domain. 

The two sources that show variability but have not previously been classified as young members nor candidate members of $\sigma$ Orionis do not have colours consistent with youth.  Therefore, these sources are not discussed any further.

\section{Searching for new members in our deep image}
\label{sec:deep_image}

We searched for new members of $\sigma$ Orionis using our constructed deep image. We used two different techniques that are detailed below.

\subsection{Searching for previously missed members using UKIDSS/GCS}
\label{subsec:xmatch}

We extracted all sources in each field of view from our final deep, stacked images using the same process as described in Section~\ref{subsec:time_series_phot}. We then cross-matched these sources with \textit{J}-band photometry from the UKIDSS/DR9 GCS catalogue \citep{Warren2007}. We chose the $J$-band as it maximises the detection efficiency and is also close to the peak of the SED for these cool objects. To identify new young candidate members, we cross-matched all of these sources with the confirmed young members and photometric candidate members of \cite{PenaRamirez2012}.  This allowed us to see the typical colour magnitude space in which young sources usually sit using our uncalibrated photometry.  Then with these known young members we constructed a lower envelope  (shown by the left grey line in Figure~\ref{fig:colour_mag}) and selected sources above this lower envelope as potential candidates. The approximate depth from our uncalibrated {\it I} -band photometry ($5\sigma\approx22.5$\,mag) combined with that of UKIDSS/DR9 GCS\footnote{\url{http://wsa.roe.ac.uk/dr9plus_release.html}.} ({\it J}$\approx$19.6\,mag) is shown by the dashed grey line in Figure~\ref{fig:colour_mag}. 

We visually inspected our images with the UKIDSS catalogue to look for any false positives producing apparently large colour differences.  For example, in some cases a bright source has a nearby ($\approx$1\,\arcsec) fainter neighbouring source in our images.  However, in the UKIDSS catalogue only the brighter source has been recovered.  This results in a large $I - J$ value which is not physical.  We initially identified 27 sources for further visual inspection; of these 27, we identified five as potential new low-mass members. One of these five has $J - K = 0.5$\,mag, which is too blue for a very low-mass member of this cluster.   Of the remaining four sources, all have been flagged as galaxies in the UKIDSS catalogue with a probability $\geq$90\%.  Table~\ref{tab:new_cands} shows details of the sources. We therefore do not present any new low-mass members using this technique. The lack of new members down to $J$ = 19\,mag, corresponding to masses just below 10\, M$_\mathrm{Jup}$, indicates that the current census as presented by \cite{PenaRamirez2012} is complete in the regions covered by our survey.

Two of the lowest mass members (S Ori 62 and S Ori 65, \textit{J}\,$>$\,19.0\,mag) are not displayed in Figure~\ref{fig:colour_mag} as there is no available UKIDSS magnitude for these two objects.  The ten young members or photometric candidates shown in Figure~\ref{fig:colour_mag} make up the remaining objects listed in Table~\ref{tab:young_sources}.

{
\begin{table}
\footnotesize
\caption{Discarded photometric cluster member candidates.}
\begin{tabular}{p{1.2cm} p{1.5cm} p{0.6cm} p{1.4cm} p{1.7cm}}
\hline  \hline\\[-1.6ex]
  \multicolumn{1}{l}{RA} &
  \multicolumn{1}{l}{DEC} &    
  \multicolumn{1}{l}{Arb. \it{I}} &
    \multicolumn{1}{l}{\it{J}} &
    \multicolumn{1}{l}{Comments}  \\
  \multicolumn{1}{l}{} &
  \multicolumn{1}{l}{} &     
  \multicolumn{1}{l}{(mag)} &    
  \multicolumn{1}{l}{(mag)} &      
  \multicolumn{1}{l}{}\\
  \hline\\
05:39:10.4 & -02:35:04 & 21.29 & 19.12$\pm$0.13 & Galaxy \\
05:38:27.3 & -02:31:26 & 20.68 & 18.61$\pm$0.09 & Galaxy\\
05:38:27.8 & -02:22:31 &21.52  & 19.40$\pm$0.19  & Galaxy\\
05:38:52.2 & -02:33:15 & 19.46 & 17.85$\pm$0.04 & Galaxy \\
05:39:20.0 & -02:27:07 & 16.95 & 15.96$\pm$0.01 & Star, {\it J-K}=0.5 \\
05:39:08.4 & -02:32:24 & \ldots & 13.79$\pm$0.01 & {\it J-K}=0.9 \\
05:38:25.8 & -02:23:09 & 21.37 & 20.30$\pm$0.10\tablefootmark{a} &  {\it I-J}=1.1 \\
05:38:21.2 & -02:33:34 & 16.40 & 15.31$\pm$0.01\tablefootmark{a} & {\it J-K}=0.9  \\
 \hline
\end{tabular}
\tablefoot{\tablefoottext{a}{K. Pe\~na Ramirez (priv. comm.).}}
\label{tab:new_cands}
\end{table}}

\subsection{Searching for companions around bona-fide members}
\label{sec:companions}

\begin{figure}
\begin{center}
\includegraphics[width=0.49\textwidth]{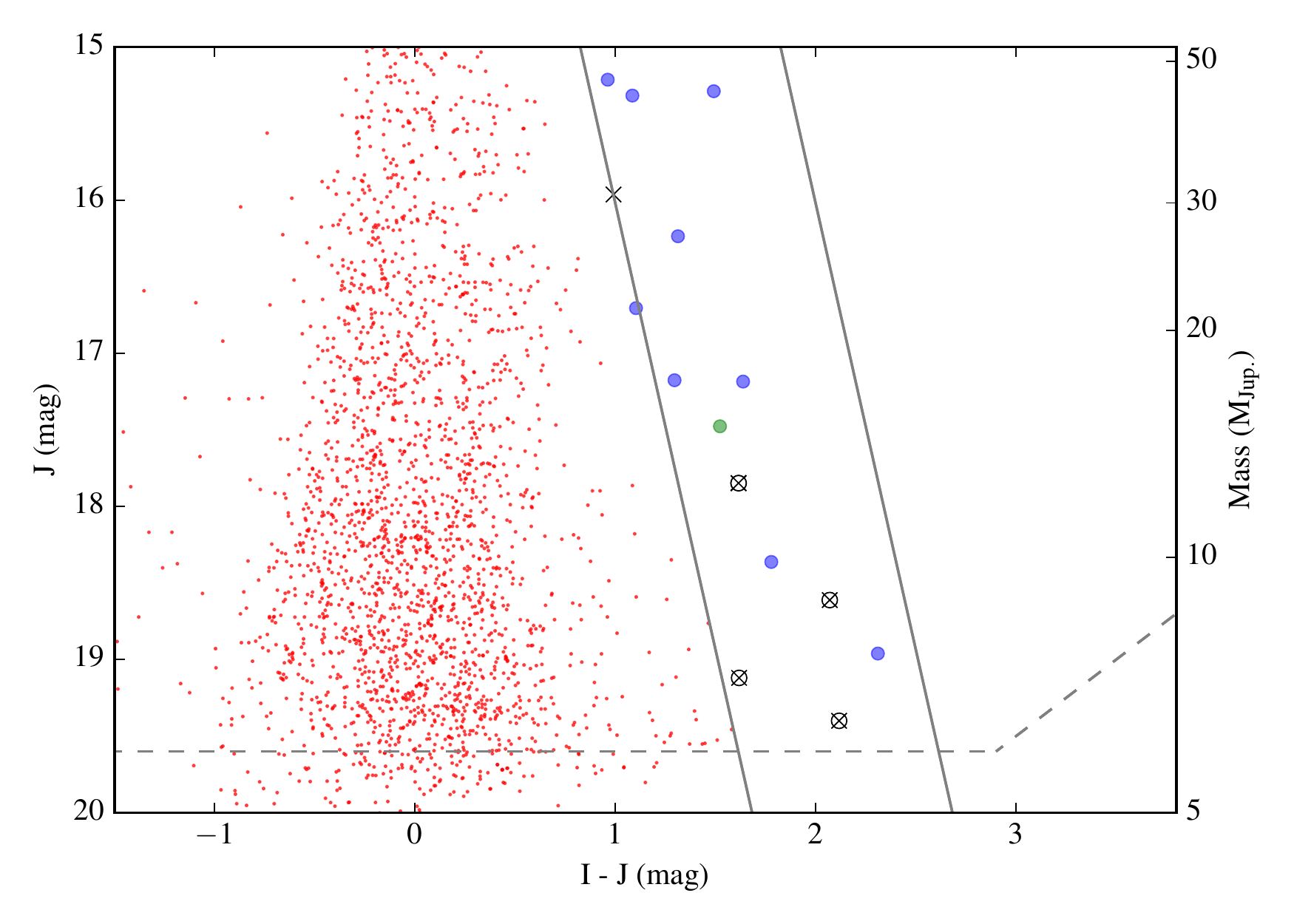}
\vspace{-0.7cm}
\caption{Colour-magnitude diagram based on UKIDSS {\it J-} and our {\it I-} band (including an offset of 31 mag) photometry. Markers are as follows, Red dots: the older field population, blue circles: young members from \cite{PenaRamirez2012}, green circles: candidate members from \cite{PenaRamirez2012}, crosses: sources initially identified as potential new members but with discrepant {\it J-K} values, crosses with circles: sources initially identified as potential new members but flagged as galaxies.  The area between the two solid grey lines defines the photometric space of potential members.  The dotted grey line is the approximate completeness limit. }
\label{fig:colour_mag}
\end{center}
\end{figure}

By searching for companions around bona-fide members we were not restricted to non-saturated sources; we were able to search around 40 young targets.  We were also able to search around the lowest-mass confirmed young members (5\,M$_\mathrm{Jup}$) for which there are no UKIDSS {\it J} magnitude values (for that reason they do not appear in Figure~\ref{fig:colour_mag}). All 40 targets that are surveyed from companions can be found in Table~\ref{tab:searched_for_comps}.

\begin{figure}
\begin{center}
\includegraphics[width=0.45\textwidth]{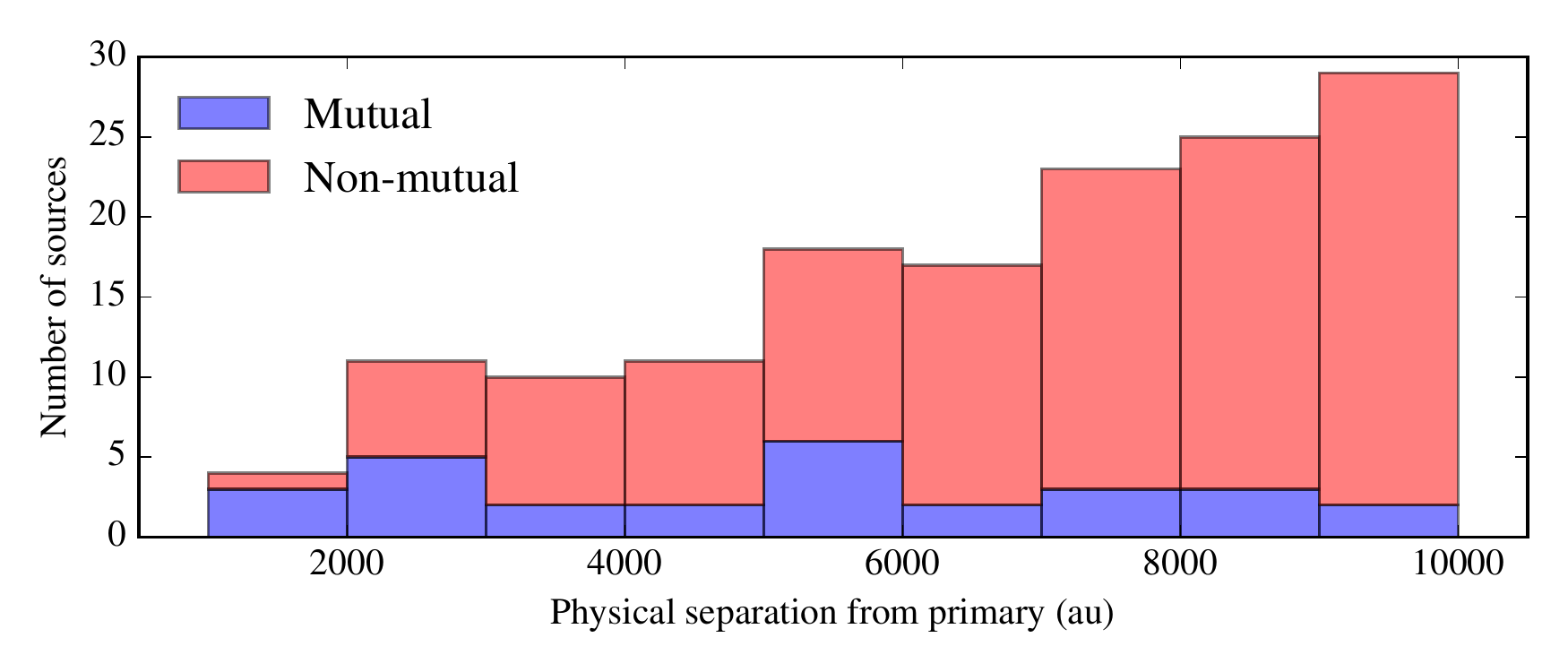}
\caption{Histogram showing the project physical separations of potential companions around all 40 bona-fide young members of $\sigma$ Orionis within 10,000\,au. A pair of sources are classified as mutual if they are each other's next nearest neighbours.}
\label{fig:mutual_neighbours}
\end{center}
\end{figure}

We determined an approximate regime in which the rate of false positives should be relatively low.  We used mutual and non-mutual nearest neighbours as a proxy for false positives.  This technique is commonly used to identify bound multiple systems in simulations \citep{Kouwenhoven2010}. A pair of sources is defined as a mutual pair if each one is the nearest neighbour of the other source.  For each potential companion, we calculated its projected physical separation and flagged whether its nearest neighbour was either the bona-fide young source or another nearby source.  We conducted this analysis for all 40 sources, searching for companions within 10,000\,au ($\approx$28\arcsec). We used this approach rather than using available colour information to exclude certain candidates as not all potential companions have counterpart infrared photometry.

Figure~\ref{fig:mutual_neighbours} shows a histogram of the mutual and non-mutual nearest neighbours as a function of physical separation for all 40 sources.  The only regime in which mutual nearest neighbours dominate over non-mutual nearest neighbours is $\leq$2000\,au.  We therefore set 2000\,au as our upper search limit on companions, wider than the limit (1550\,au) used in  \cite{Caballero2017b}. Our inner search limit is determined by the size of the Point Spread Function (PSF), $\approx$1.5\arcsec ($\approx$500\,au).   Three new sources were identified as mutual nearest neighbours within this physical separation.  

The median mass of the 40 objects that we considered was 0.09\,M$_\odot$.  Therefore, it is not entirely surprising that we did not discover any new companions given previous studies of multiplicity in similar mass regimes.  The multiplicity fraction and peak of the physical separation distribution are a strong function of primary mass (see Figure~1 of \citealt{Duchene2013}).   \cite{Burgasser2006}, \cite{Ahmic2007}, \cite{Luhman2012}, and \cite{Janson2014}, to cite a few examples, have studied the multiplicity of low-mass stars and brown dwarfs in both young regions and the field.  In all cases, companions to low-mass primaries beyond 100\,au are extremely rare ($<$1\,\%).  The peak and standard deviation of the physical separation distribution for late M dwarfs is $\approx$6$^{+13}_{-4}$\,au \citep{Janson2014}.  In other words, beyond our minimum separation range ($\approx$500\,au) even for the highest mass primaries (best case scenario for companion detections) of our sample, we are already beyond 5\,$\sigma$ from the mean of the expected distribution.

Additionally we checked the list of 40 sources with a recent multiplicity review of $\sigma$ Orionis \citep{Caballero2014} in case we had missed any previously documented detections.  We did not find any evidence of multiplicity for these targets within our angular separation limit (1.5-5.7\arcsec).

\section{Conclusions}
\label{sec:conclusions}

In this work we have presented two main techniques in the identification and study of substellar objects in $\sigma$ Orionis using VIMOS/VLT {\it I}-band observations. Below are the main findings and conclusions from our analysis.

\begin{itemize}
\item We have identified significant variability in two young brown dwarfs, one newly identified, from a sample of nine.
\item Given the short time span of observations and the strong inter-night variations in their quasi-periodic signal, we could not calculate a definitive period for either object.
\item The first object, Mayrit 258337 (a single-lined spectroscopic binary), shows a host of consistent properties with other young variable objects, such as correlated and variable mid-infrared magnitudes and mid-infrared excess. Therefore, its variability in the $I$ band is most likely linked to its accretion disc.
\item The second object, Mayrit 396273, has no mid-infrared excess and no significant correlation or variation in mid-infrared magnitude.  The observed variability in the $I$ band may be caused by variable extinction by large grains.
\item We did not find any new low-mass potential members of $\sigma$ Orionis using our uncalibrated {\it I}-band photometry with available UKIDSS $J$-band photometry, consistent with the results of \cite{PenaRamirez2012}.
\item We did not identify any new low-mass companions around forty young $\sigma$ Orionis sources in the approximate physical separation range 500-2000\,au, consistent with other studies of wide multiplicity in very low-mass objects.
\end{itemize}

Our uncalibrated {\it I}-band photometry for sources successfully cross-matched with UKIDSS/DR9 GCS catalogue \citep{Warren2007} is available publicly via the VizieR service.

\begin{acknowledgements}
We  thank  the  referee,  Jos\'e A.  Caballero,  for  constructive
comments.  This work has made use of data from the European Space Agency (ESA) mission {\it Gaia} (\url{http://www.cosmos.esa.int/gaia}), processed by
the {\it Gaia} Data Processing and Analysis Consortium (DPAC,
\url{http://www.cosmos.esa.int/web/gaia/dpac/consortium}). Funding
for the DPAC has been provided by national institutions, in particular
the institutions participating in the {\it Gaia} Multilateral Agreement. Additionally, we have used data from the UKIDSS project, defined in \cite{Lawrence2007}. UKIDSS uses the UKIRT Wide Field Camera (WFCAM; \citealt{Casali2007}). The photometric system is described in \cite{Hewett2006}, and the calibration is described in \cite{Hodgkin2009}. The pipeline processing and science archive are described in \cite{Hambly2008}.    This research has also made extensive use of the Topcat tool \citep{Taylor2005}, the SIMBAD database, and the VizieR catalogue access tool, CDS, Strasbourg, France (the original description of the VizieR service was published in A\&AS, 143, 23).  This publication makes use of data products from the Wide-field Infrared Survey Explorer, which is a joint project of the University of California, Los Angeles, and the Jet Propulsion Laboratory/California Institute of Technology, and NEOWISE, which is a project of the Jet Propulsion Laboratory/California Institute of Technology. WISE and NEOWISE are funded by the National Aeronautics and Space Administration.  Figure~\ref{fig:2mass_pointings} was made using {\sc APL}py, an open-source plotting package for Python \citep{Robitaille2012}.

\end{acknowledgements}

\bibliography{./biblio1}

\Online

\begin{appendix}

{
\onecolumn
\section{Young sources used in the search for wide, faint companions}
\label{sec:appendix}

\LTcapwidth=\textwidth
\begin{longtable}{p{6cm} p{2cm} p{2cm} p{1.4cm} p{1.7cm}}
\caption{$\sigma$ Orionis stars and brown dwarfs that we searched for close companions.}
\label{tab:searched_for_comps} \\
\hline  \hline\\[-1.6ex]
  \multicolumn{1}{l}{Simbad ID} &
  \multicolumn{1}{l}{RA} &    
  \multicolumn{1}{l}{DEC} &
    \multicolumn{1}{l}{\it{J}} &
    \multicolumn{1}{l}{$\mathcal{M}$\,\tablefootmark{a}}  \\
  \multicolumn{1}{l}{} &
  \multicolumn{1}{l}{hh:mm:ss.ss} &     
  \multicolumn{1}{l}{dd:mm:ss.s} &    
  \multicolumn{1}{l}{(mag)} &      
  \multicolumn{1}{l}{(M$_\odot$)}\\
  \hline\\
{[BZR99]}\,S Ori 13                  &  05 38 13.21  &  -02 24 07.6  &  14.06  &  0.156 \\
{[BZR99]}\,S Ori 9                   &  05 38 17.18  &  -02 22 25.7  &  13.55  &  0.233 \\
{[BMZ2001]}\,S Ori J053818.2-023539  &  05 38 18.34  &  -02 35 38.5  &  15.29  &  0.054 \\
{[BMZ2001]}\,S Ori J053821.3-023336  &  05 38 21.38  &  -02 33 36.2  &  15.31  &  0.053 \\
{[SWW2004]}\,J053822.999-023649.48    &  05 38 23.07  &  -02 36 49.4  &  13.73  &  0.200 \\
{[SWW2004]}\,J053823.351-022534.51    &  05 38 23.34  &  -02 25 34.6  &  13.66  &  0.212 \\
{[BZR99]}\,S Ori 18                  &  05 38 25.68  &  -02 31 21.6  &  14.59  &  0.093 \\
{[BZR99]}\,S Ori 65                  &  05 38 26.10  &  -02 23 05.0  &  20.30  &  0.005 \\
{[BZR99]}\,S Ori 29                  &  05 38 29.62  &  -02 25 14.2  &  14.79  &  0.077 \\
{[FPS2006]}\,NX 46                    &  05 38 32.13  &  -02 32 43.1  &  13.16  &  0.313 \\
{[BZR99]}\,S Ori 22                  &  05 38 35.35  &  -02 25 22.2  &  14.59  &  0.093 \\
{[HHM2007]}\,633                      &  05 38 38.12  &  -02 32 02.6  &  15.06  &  0.053 \\
{[KJN2005]}\,65                       &  05 38 39.76  &  -02 32 20.3  &  14.94  &  0.074 \\
{[BZR99]}\,S Ori 6                   &  05 38 47.66  &  -02 30 37.4  &  13.39  &  0.263 \\
{[BZR99]}\,S Ori 15                  &  05 38 48.10  &  -02 28 53.6  &  14.37  &  0.116 \\
{[KJN2005]}\,8                        &  05 38 50.78  &  -02 36 26.7  &  13.06  &  0.337 \\
{[MJO2008]}\,J053852.6-023215         &  05 38 52.63  &  -02 32 15.5  &  16.18  &  0.028 \\
{[SWW2004]}\,J053854.916-022858.24    &  05 38 54.93  &  -02 28 58.3  &  13.72  &  0.200 \\
{[BZR99]}\,S Ori 71                  &  05 39 00.30  &  -02 37 05.8  &  17.19  &  0.018 \\
{[KJN2005]}\,9                        &  05 39 01.16  &  -02 36 38.8  &  13.51  &  0.242 \\
{[BMZ2001]}\,S Ori J053902.1-023501  &  05 39 01.94  &  -02 35 02.9  &  14.74  &  0.075 \\
{[BZR99]}\,S Ori 51                  &  05 39 03.20  &  -02 30 20.0  &  17.16  &  0.018 \\
{[BZR99]}\,S Ori 58                  &  05 39 03.60  &  -02 25 36.0  &  18.42  &  0.011 \\
{[BZR99]}\,S Ori 17                  &  05 39 04.49  &  -02 38 35.3  &  14.71  &  0.080 \\
{[BZR99]}\,S Ori 20                  &  05 39 07.61  &  -02 29 05.7  &  14.90  &  0.076 \\
{[BZR99]}\,S Ori 8                   &  05 39 08.09  &  -02 28 44.8  &  14.07  &  0.155 \\
{[BZR99]}\,S Ori 7                   &  05 39 08.22  &  -02 32 28.4  &  13.77  &  0.194 \\
{[BMZ2001]}\,S Ori J053911.4-023333  &  05 39 11.40  &  -02 33 32.8  &  14.41  &  0.100 \\
{[BMZ2001]}\,S Ori J053912.8-022453  &  05 39 12.89  &  -02 24 53.5  &  16.68  &  0.022 \\
{[BZR99]}\,S Ori 30                  &  05 39 13.08  &  -02 37 50.9  &  15.20  &  0.059 \\
{[HHM2007]}\,1075                     &  05 39 29.35  &  -02 27 21.0  &  13.10  &  0.326 \\
{[BZR99]}\,S Ori 21                  &  05 39 34.33  &  -02 38 46.9  &  14.69  &  0.090 \\
{[BZR99]}\,S Ori 60                  &  05 39 37.50  &  -02 30 42.0  &  19.02  &  0.008 \\
{[BZR99]}\,S Ori 62                  &  05 39 42.05  &  -02 30 31.6  &  19.14  &  0.008 \\
{[BMZ2001]}\,S Ori J053950.6-023414  &  05 39 50.56  &  -02 34 13.7  &  13.62  &  0.219 \\
{[BMZ2001]}\,S Ori J053954.2-022733  &  05 39 54.20  &  -02 27 32.7  &  13.45  &  0.252 \\
{[BMZ2001]}\,S Ori J053954.3-023720  &  05 39 54.32  &  -02 37 18.9  &  14.69  &  0.090 \\
{[BMZ2001]}\,S Ori J053956.4-023804  &  05 39 56.45  &  -02 38 03.5  &  13.30  &  0.281 \\
 \hline\\
\end{longtable}
\tablefoot{\tablefoottext{a}{Using {\it J} magnitude, a distance of 352\,pc, and an age of 5\,Myr for the evolutionary models of \cite{Baraffe2003a} and \cite{Baraffe2015}}.}
}
\newpage
{

\section{Photometric catalogue}

\begin{table*}
\begin{center}
\caption{First 20 entries of uncalibrated \textit{I}-band photometry for cross-matched UKIDSS sources.}
\begin{tabular}{p{2.5cm} p{1.8cm} p{1.8cm} p{1.5cm} p{2cm} p{2cm} p{1.5cm}}
\hline\hline\\[-1.6ex]
  \multicolumn{1}{l}{UKIDSS ID} &
  \multicolumn{1}{l}{RA} &
  \multicolumn{1}{l}{DEC} &  
  \multicolumn{1}{l}{$J$} &
  \multicolumn{1}{l}{$\sigma_J$} &
  \multicolumn{1}{l}{$I_{\,\mathrm{arb}}$} &
  \multicolumn{1}{l}{Field + Quadrant} \\
    \multicolumn{1}{l}{} &
  \multicolumn{1}{l}{hh:mm:ss.ss} &
  \multicolumn{1}{l}{dd:mm:ss.s} &    
  \multicolumn{1}{l}{(mag)} &
  \multicolumn{1}{l}{(mag)} &
  \multicolumn{1}{l}{(mag)} &
  \multicolumn{1}{l}{} \\
\hline\\
  442414367679 & 05:39:01.38 & -02:28:18.5 & 18.26 & 0.06 & 18.05 & A2, B3\\
  442414367681 & 05:39:14.04 & -02:28:17.7 & 17.24 & 0.03 & 17.12 & A2\\
  442414367682 & 05:39:27.83 & -02:28:16.7 & 17.92 & 0.05 & 17.58 & B2\\
  442414367684 & 05:39:49.02 & -02:28:14.2 & 16.45 & 0.01 & 16.11 & B2\\
  442414367686 & 05:39:17.35 & -02:28:12.8 & 16.92 & 0.02 & 15.96 & A2\\
  442414367688 & 05:39:10.02 & -02:28:11.5 & 14.51 & 0.0 & 15.08 & A2\\
  442414367689 & 05:39:09.22 & -02:28:09.9 & 18.29 & 0.07 & 17.99 & B3\\
  442414367694 & 05:39:47.36 & -02:28:08.0 & 18.83 & 0.11 & 19.21 & B2\\
  442414367707 & 05:39:07.33 & -02:28:01.2 & 18.63 & 0.09 & 18.95 & B3\\
  442414367708 & 05:38:59.66 & -02:28:00.5 & 18.08 & 0.06 & 18.19 & B3\\
  442414367709 & 05:39:51.79 & -02:28:00.3 & 17.44 & 0.03 & 17.33 & B2\\
  442414367714 & 05:39:51.22 & -02:27:59.0 & 17.52 & 0.03 & 17.82 & B2\\
  442414367716 & 05:39:04.11 & -02:27:57.8 & 17.97 & 0.05 & 17.78 & B3\\
  442414367717 & 05:39:02.05 & -02:27:58.1 & 16.83 & 0.02 & 17.41 & A2, B3\\
  442414367721 & 05:39:43.93 & -02:27:55.2 & 19.04 & 0.13 & 18.53 & B2\\
  442414367732 & 05:39:45.26 & -02:27:50.1 & 18.79 & 0.1 & 18.45 & B2\\
  442414367733 & 05:39:37.84 & -02:27:49.5 & 19.24 & 0.16 & 19.51 & B2\\
  442414367739 & 05:38:59.36 & -02:27:47.0 & 18.18 & 0.06 & 17.66 & A2, B3\\
  442414367740 & 05:39:37.71 & -02:27:46.1 & 18.08 & 0.06 & 17.53 & B2\\
  442414367743 & 05:39:06.43 & -02:27:45.0 & 15.53 & 0.01 & 16.07 & A2, B3\\
  442414367747 & 05:39:03.76 & -02:27:41.5 & 18.56 & 0.08 & 18.81 & A2, B3\\
  \hline
\end{tabular}
\label{tab:ukidss_vimos_cat}
\end{center}
\end{table*}
}
\end{appendix}

\end{document}